\documentstyle[prl,aps,floats,epsf]{revtex}\def\narrowtext{}\tighten\twocolumn
\input epsf.sty
\begin{document}
\draft

\title{
Zn-Neighbor Cu NQR in Zn-Substituted  
YBa$_2$Cu$_3$O$_{7-\delta}$ and YBa$_2$Cu$_4$O$_8$
}
\author{
Y. Itoh,$^{1,2}$ T. Machi,$^1$ C. Kasai,$^1$ S. Adachi,$^1$ N. Watanabe,$^1$ N. Koshizuka,$^1$ and M. Murakami$^1$ 
}

\address{
$^1$Superconductivity Research Laboratory,
 International Superconductivity Technology Center,\\
1-10-13 Shinonome, Koto-ku Tokyo 135-0062, Japan \\
$^2$Japan Society for the Promotion of Science, Tokyo 102-8471, Japan\\
}

\date{\today}%
\maketitle %

\begin{abstract}
We studied local electronic states near Zn in optimally doped 
YBa$_2$(Cu$_{1-x}$Zn$_x$)$_3$O$_{7-\delta}$ 
and underdoped YBa$_2$(Cu$_{1-x}$Zn$_x$)$_4$O$_8$  
via satellite signals of 
plane-site Cu(2) nuclear quadrupole resonance (NQR) spectra. 
From the relative intensity of Cu NQR spectra, 
the satellite signals are assigned to Zn-neighbor Cu NQR lines.
The Cu nuclear spin-lattice relaxation time of the satellite signal is shorter than 
that of the main signal, 
which indicates that the magnetic correlation is locally enhanced near Zn 
both for the underdoped and the optimally doped systems. 
The pure YBa$_2$Cu$_4$O$_8$ is a stoichiometric, homogenous, underdoped electronic system;  
nevertheless, the Zn-induced inhomogeneous magnetic response in the CuO$_2$ plane   
is more marked than that of the optimally doped YBa$_2$Cu$_3$O$_{7-\delta}$. 
\end{abstract}
\pacs{76.60.-k, 74.25.Nf, 74.72.Bk}

\narrowtext

\section{Introduction}
\label{sec:intro}

The effect of nonmagnetic impurity Zn 
in high-temperature cuprate superconductors has attracted great interests. 
The impurity Zn induces Curie magnetism, a local staggered susceptibility
in the normal state~\cite{Xiao,Miyatake,Mahajan,Walstedt,Julien},
an in-gap antiferromagetic dynamical spin susceptibility, and 
a virtual bound electronic state around Zn in the
superconducting state~\cite{Sidis,Pan,Itoh}.  
The perturbation theory
for the mean-field states close to two-dimensional antiferromagnetic instability 
allows coexistence of the staggered-type, generalized 
Ruderman-Kittel-Kasuya-Yosida (RKKY) oscillation and the local density of states of electrons with a zero bias resonance peak
at finite temperatures~\cite{Kilian,Ohashi,Ohashi1}. 
However, recent numerical calculation using G\"{u}tzwiller approximation
indicates that the frozen staggered moments are locally reduced by Zn in the
superconducing state of a moderately doped two-dimensional $t-J$ model~\cite{Tsuchiura}.
According to this theory, the zero bias resonance peak in the local density of states 
and the local staggered moments are exclusive at $T$=0 K.
It remains to be solved whether the Zn
strengthens~\cite{Yamagata,AlloulBob} or weakens~\cite{Ishida} local magnetic correlation around Zn 
in the optimally doped or overdoped system.     
 
Nuclear magnetic resonance (NMR) and nuclear quadrupole resonance (NQR) 
are powerful techniques to study the local states 
on crystalline imperfection, dilute exchange spin networks, Friedel oscillation, RKKY
oscillation, or Kondo screening effect
~\cite{Red,Jac,Berthier,Boyce,Alloul}.  
We focus on the local electronic states near Zn in YBa$_2$(Cu$_{1-x}$Zn$_x$)$_3$O$_y$ ($y\equiv$7-$\delta$) (Y123)
 and YBa$_2$(Cu$_{1-x}$Zn$_x$)$_4$O$_8$ (Y124)
via Zn-induced Cu NQR satellite spectra~\cite{Yamagata,WilliamsCu}. 
Y123 is a nearly optimally doped system (oxygen content $y\sim$6.9), 
while Y124 is a stoichiometric underdoped one. 
Zn-neighbor $^{89}$Y NMR signals and their relaxation times have intensively been studied 
for Y123~\cite{Mahajan} and for Y124~\cite{WilliamsY}.
However, there is only a few direct observation of the local electronic state via Zn-neighbor Cu NQR/NMR signals; 
a report on the satellite Cu nuclear spin-lattice relaxation 
of a non-superconducting Y124 with heavily Zn doping~\cite{WilliamsCu} 
but no report on the superconducting Y123 or Y124 with small concentrations of Zn.

In this paper, we report on a study of satellite signals in the planar Cu(2) NQR spectra 
of Zn-substituted superconductors Y123 and Y124. 
The wipeout effect on the Cu NQR main signals and the growth of the Zn-induced Cu satellite signals 
indicate that the satellite but not the main signal includes the Zn-neighbor Cu nuclei.
We demonstrate that the satellite Cu nuclear spin-lattice relaxation time is shorter than the main
one in the superconducting Y123 and Y124.  
The local antiferromagnetic correlation near Zn is enhanced even for the optimally doped system.  

\section{Experimental}
\label{sec:expt}

High quality powder samples of a pure Y123 ($y$=6.98 studied in Ref.~\cite{Shimizu0}), 
a slightly oxygen-deficient Y123 ($y$=6.89 in Refs.~\cite{Matsumura,Yamagata}), 
Zn-substituted Y123 (the annealed samples in~\cite{Adachi}), 
a pure Y124 and Zn-substituted Y124 ~\cite{Itoh,Itoh1} were prepared for the present study. 
The superconducting transition temperatures are 
$T_c$$\sim$92 K both for Y123 with $y$=6.98 and with $y$=6.89; 
$T_c$=81, 69, and 46 K for Y123 with Zn content $x$=0.007, 0.017, 0.033; 
$T_c$=82 K for the pure Y124, and $T_c$=68, 56, 37, and 27 K 
for Y124 with Zn content $x$=0.005, 0.01, 0.015, 0.022, respectively.   
To estimate the relative intensity of Cu NQR spectra, the respective samples of Y123 and of Y124
with nearly the same volume were prepared. 
The oxygen contents conform to the values in the previous studies.    

A phase-coherent-type pulsed spectrometer was utilized to perform Cu NQR experiments. 
The zero-field Cu NQR frequency spectra with quadrature detection
were obtained by integration of Cu nuclear spin-echoes with changing the frequency in a two-pulse sequence
($\pi$/2-$\tau$-$\pi$ echo). A typical width of the first exciting $\pi$/2-pulse $t_w$ was about 3 $\mu$s 
(the excited frequency region $\nu_1\sim$83 kHz from 2$\pi\nu_1t_w$=$\pi$/2).  
The Cu nuclear spin-lattice relaxation curves were measured by an inversion recovery technique, 
where the nuclear spin-echo
$M(t)$ was monitored as a function of the time $t$ in a sequence of $\pi-t-\pi/2-\tau-\pi$ echo. 
The pulse interval for the NQR spectrum measurement was $\tau$=15 $\mu$s. 

\section{Results and discussion}
\label{sec:results}
\subsection{Main and satellite $^{63, 65}$Cu(2) NQR spectrum} 
\subsubsection{Frequency-distributed Cu(2) nuclear spin-lattice relaxation}

Figure 1 shows $^{63, 65}$Cu(2) NQR spectra (upper panels) and 
the $^{63}$Cu(2) nuclear spin-lattice relaxation curves $p(t)\equiv1-M(t)/M(\infty)$ 
(lower panels) for the oxygen-deficient Y123 (left), Zn-substituted, nearly optimally doped Y123 (middle), 
and Zn-substituted, underdoped Y124 (right) at $T$=4.2 K. 
The Cu NQR spectra are normalized by the respective values of peak intensity. 
Zn doping induces lower frequency tails or satellite signals both for Y123 and Y124,
which is known in Refs.~\cite{Yamagata,WilliamsCu}. 
Oxygen deficiency is also known to yield similar low frequency satellite or shoulder in the Cu(2) NQR spectrum 
for Y123 without Zn
~\cite{Vega,Yasuoka,Warren,Imai0}. 
Since Y124 has no appreciable oxygen deficiency, the satellite signals result from purely Zn doping effect.

\begin{figure}
\epsfxsize=3.7in
\epsfbox{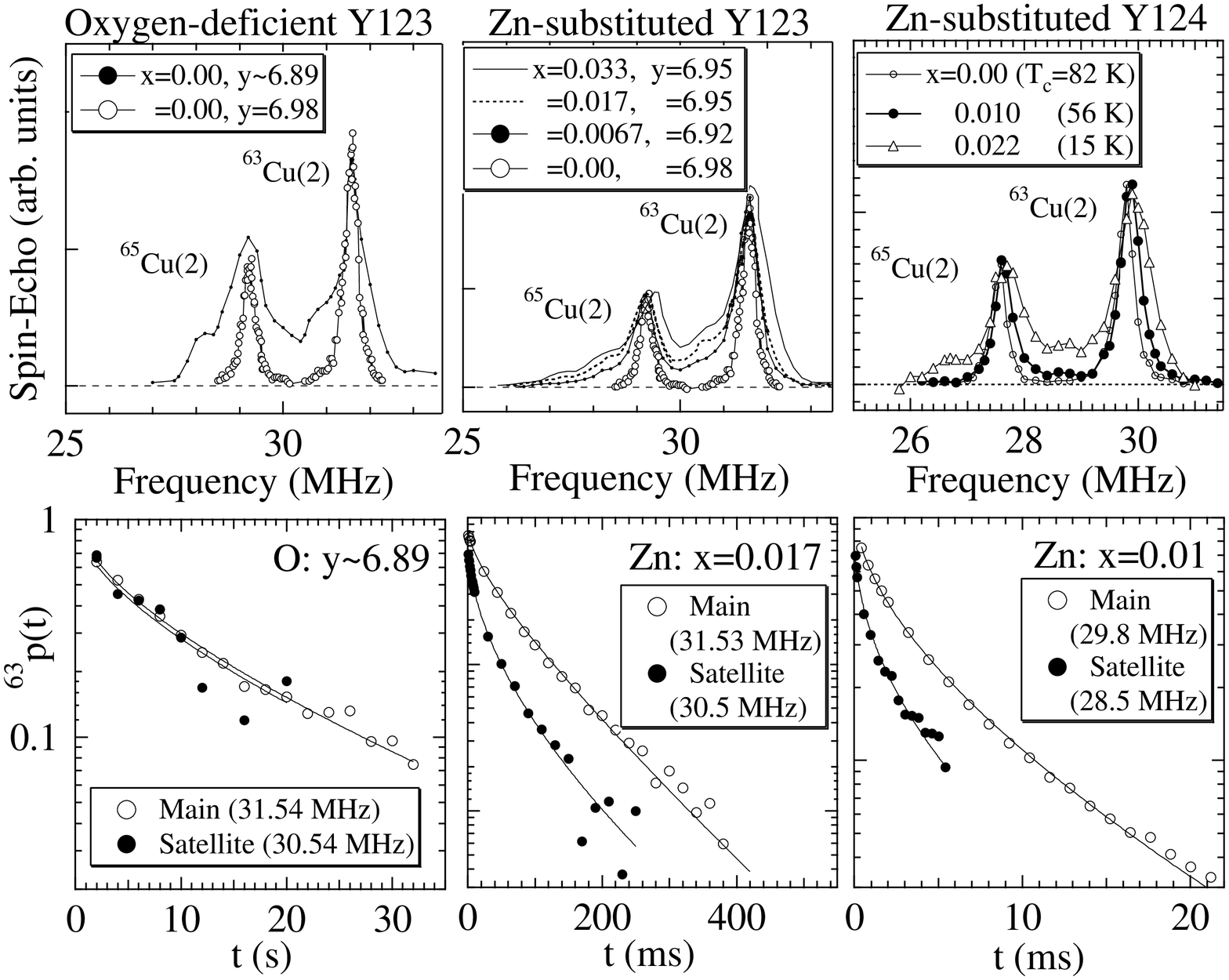}
\vspace{0.2cm}
\caption{
Upper panels: Zero-field frequency spectra of the planar $^{63, 65}$Cu(2) NQR 
for oxygen-deficient Y123 with $y$=6.89 and $y$=6.98 
(left), 
Zn-substituted Y123 (middle), and 
Zn-substituted Y124 (right) at $T$=4.2 K. 
Lower panels: The main and the satellite $^{63}$Cu(2) nuclear spin-lattice relaxation curves $p(t)$ for the representative samples at
$T$=4.2 K. 
The solid curves are fitted by theoretical recovery curves (see the text). 
For Y124 (Y123), the satellite recovery curve $p(t)$ is nearly independent of frequency 
over 28.4-28.6 MHz (30.1-30.5 MHz) 
and so is the main one over 29.2-30.2 MHz (31.5-31.6 MHz).  
}
\label{T1Distribution}
\end{figure}

For oxygen deficient Y123 in the lower panel of Fig.1, the satellite and 
the main signals show nearly the same relaxation curve $p(t)$, 
and their relaxation times $T_1$ are longer than a few seconds.  
This is consistent with the previous reports~\cite{Vega,Imai0}. 
On the other hand, for Zn-doped Y123 and Y124, the Zn-induced satellite signal shows a shorter relaxation
time than the main signal, and their relaxation times are shorter than 1.0 s. 
Thus, the lower frequency satellite in the Cu NQR spectrum for Zn-doped Y123 possesses an electronic state different from 
that for the slightly oxygen deficient one.

Our observation for Zn-substituted Y123 and Y124 that 
the lower frequency signals in the broad NQR spectra show the shorter $T_1$ is 
in parallel to that for Zn-free underdoped La$_{2-x}$Sr$_x$CuO$_4$~\cite{Fujiyama,Imai}. 
For Y124 (Y123), we confirm that the satellite recovery curve $p(t)$ is nearly independent of frequency 
over 28.4-28.6 MHz (30.1-30.5 MHz) 
and so is the main one over 29.2-30.2 MHz (31.5-31.6 MHz).
Here, the excited frequency region by the pulse is $\nu_1\sim$83 kHz at each NQR frequency. 
Thus, the frequency dependence of the relaxation curve is not monotonic. 

\subsubsection{Isotope dependence of Cu(2) nuclear spin-lattice relaxation} 

Figure 2(a) shows the $^{63, 65}$Cu(2) NQR spectrum of Y123 with Zn content of $x$=0.033 ($T_c$=46 K) at $T$=4.2 K 
and the hot spot regions (shaded bars) excited in this spectrum 
to measure the isotope dependence of Cu(2) nuclear spin-lattice relaxation curves.
Figure 2(b) shows the respective $^{63, 65}$Cu(2) nuclear spin-lattice relaxation curves $p(t)$ 
for the above Y123 at $T$=4.2 K.
The dashed and the solid curves are the respective least-squares fitting results 
for $^{63}$Cu(2) and $^{65}$Cu(2), 
using a theoretical function including a stretched exponential function (see below).
The nuclear spin-lattice relaxation rate due to a magnetic origin is proportional  
to the square of the $^{63, 65}$Cu nuclear gyromagnetic ratio
$(^{63, 65}\gamma_n)^2$. 
Since we could not find any large deviation from the scaling with respect to the reduced time $(^{63, 65}\gamma_n)^2t$, 
the overlapping effect of the broad $^{65, 63}$Cu(2) lines is small.

\begin{figure}
\epsfxsize=3.5in
\epsfbox{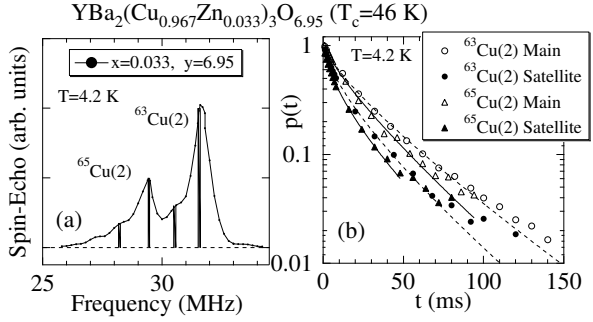}
\vspace{0.1cm}
\caption{
(a) The $^{63, 65}$Cu(2) NQR spectrum of Y123 with Zn $x$=0.033 ($T_c$=46 K) at $T$=4.2 K 
and the hot spot regions (shaded bars) excited by rf pulses to measure the nuclear spin-lattice relaxation.
(b) The $^{63, 65}$Cu(2) nuclear spin-lattice relaxation curves $p(t)$ measured at the respective frequencies denoted in (a). 
The dashed ($^{63}$Cu(2)) and the solid ($^{65}$Cu(2)) curves are the least-squares
fitting results using a theoretical function including a stretched exponential function (see the text).   
}
\label{Recovery63&65}
\end{figure}

\subsection{Zn-substitution effect on $^{63}$Cu(2) NQR spectra}
\subsubsection{Zn-induced wipeout effect and satellite}

Figure 3 shows the $^{63}$Cu(2) NQR spectra with Zn doping at $T$=100 K($>$ $T_c$)  
for Y123 with Zn contents of $x$=0, 0.007, 0.017, 0.033 (a), 
and for Y124 with Zn $x$=0, 0.01, 0.015, 0.022 (b) from the top to the bottom. 
The normal-state Cu NQR intensity is not affected by the Meissner shielding effect. 
The main signal intensity of the $^{63}$Cu(2) NQR spectrum around 31.5 MHz for Y123 and around 29.8 MHz for Y124 
rapidly decreases with increasing Zn concentration. 
Alternatively, the lower frequency tail or the broad satellite signal increases with increasing Zn concentration.
The solid and the dotted curves are numerical simulations using three Gaussian functions ($M_1$, $M_2$, and $S$). 
By using two Gaussian functions, $M_1$ and $M_2$, we can reproduce the main spectra. 
The shaded Gaussian $S$ is the satellite spectrum. 

\begin{figure}
\epsfxsize=3.7in
\epsfbox{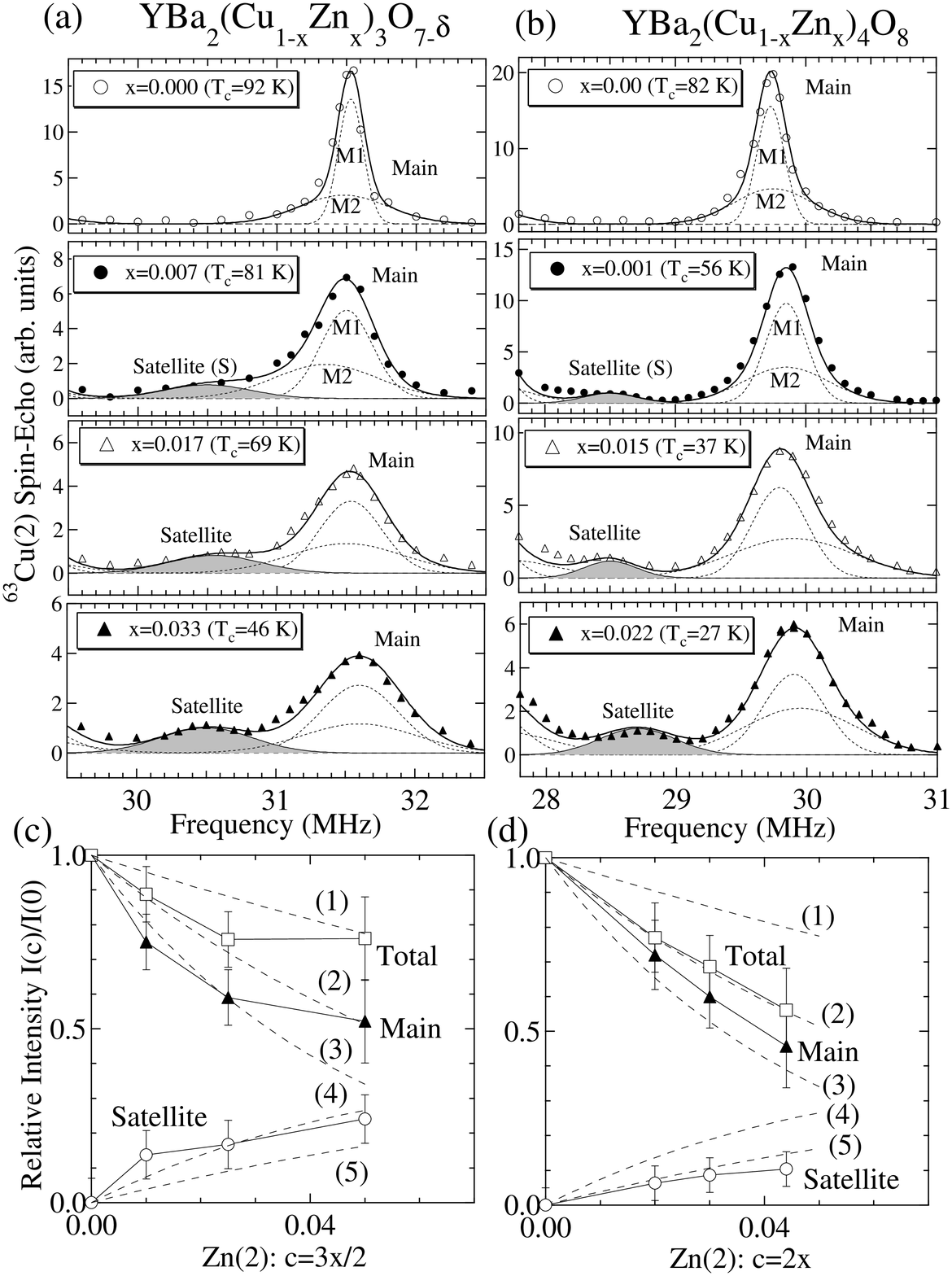}
\vspace{0.0cm}
\caption{
Zn-doping effect on the plane-site 
$^{63}$Cu(2) NQR  spectra for Y123 (a) and for Y124 (b) at $\tau$=15 $\mu$s and at $T$=100 K ($>$ $T_c$).
The line shapes are scaled before $T_2$ corrections. The relative intensity
$I_{main}/I(0)$ (upward closed triangles), 
$I_{sate}/I(0)$ (open circles), 
and the total intensity (open squares)
estimated after $T_2$ corrections 
are plotted as functions of the in-plane Zn(2) concentration $c$(=3$x$/2) for Y123 (c) 
and (=2$x$) for Y124 (d). 
The dashed curves in (c) and (d) are $I(c)/I(0)$=$(1-c)^{5}$(1), $(1-c)^{13}$(2), 
$(1-c)^{21}$(3), $8c(1-c)^{8}$(4), and $4c(1-c)^{4}$(5).
}
\label{CuNQRZn}
\end{figure}

\subsubsection{Relative intensity}

In Figs. 3(c) and 3(d), 
the estimated intensity of the main spectrum $I_{main}$(=$M_1$+$M_2$) (upward closed triangles) and 
that of the satellite spectrum $I_{sate}$(=$S$) (open circles)
normalized by $I(0)$ are plotted as functions of the plane-site Zn(2) concentrations $c$=3$x$/2 for Y123
(c) and of $c$=2$x$ for Y124 (d). 
The open squares are the total intensity(=$M1$+$M2$+$S$). 
The values of $I_{main}$ and $I_{sate}$ are estimated after $T_2$ correction using $M(0)$ from fitting
$M(\tau \geq 15 \mu$s)=$M(0)$exp$[-2\tau/T_{2L}-0.5(2\tau/T_{2G})^2]$ 
with the fitting parameters $M(0)$, $T_{2L}$ and $T_{2G}$.   

The dashed curves are numerical calculations of Cu intensity with various Zn configurations, 
using the binomial-probability 
$_{n(j)}B_k(c)$$\equiv$$_{n(j)}C_kc^k(1-c)^{n(j)-k}$, 
where $n(j)$ is the number of the $j$th nearest neighbor (nn) Cu sites.  
(1-$c$) is the probability that a plane site is occupied 
by a Cu atom but not a Zn atom, or it is the total Cu NQR intensity with Zn doping.
With respect to Zn configuration, the probability of finding Cu atoms 
can be seen in the decomposition of (1-$c$)=(1-$c$)$\prod_{j=1}\sum_{k=0}^{n(j)}$$_{n(j)}B_k(c)$. 
In Fig. 4, schematic illustrations of a $^{63}$Cu(2) NQR spectrum (a) and 
the CuO$_2$ plane with a Zn impurity (b) are given for a guide to the calculation.

\begin{figure}
\epsfxsize=3.5in
\epsfbox{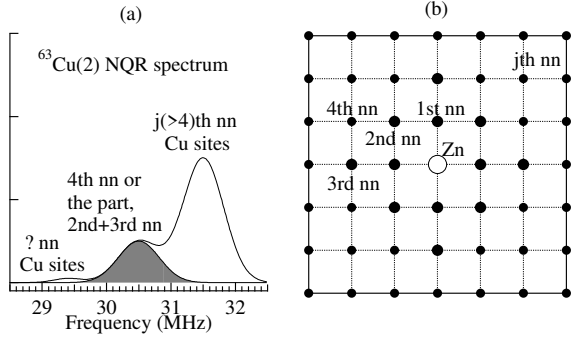}
\vspace{0.0cm}
\caption{
Schematic illustrations of (a) a $^{63}$Cu(2) NQR spectrum and (b) a top view of the CuO$_2$ plane with a Zn impurity.
The shaded area is assigned to the Zn-neighbor Cu sites. 
}
\label{CuO2}
\end{figure}

The calculated curves on the Cu NQR intensity are in what follows.  
In Figs. 3(c) and 3(d), the decreasing functions with increasing $c$ are $I(c)/I(0)$=
$(1-c)_4B_0(c)$=$(1-c)^5$ (1), 
$(1-c)(_4B_0(c))^3$=$(1-c)^{13}$ (2), and 
$(1-c)(_4B_0(c))^4$=$(1-c)^{21}$ (3). 
These are the probability of finding the Cu atoms with all Cu atoms in their 1st nn positions (1), 
in their 12 neighboring (up-to 3rd nn) positions (2), and in their 20 neighboring (up-to 4th nn) positions (3). 
In other words, the diminished Cu NQR signals come from the Cu atoms with at least one Zn atom 
in their 1st nn positions (the 1st nn wipeout effect) (1), 
in their up-to 3rd nn positions (up-to 3rd nn wipeout effect) (2), and 
in their up-to 4th nn positions (up-to 4th nn wipeout effect) (3). 
In Figs. 3(c) and 3(d), the increasing functions with increasing $c$ are $I(c)/I(0)$=   
$(1-c)_8B_1(c)$=$8c(1-c)^8$ (4), and
$(1-c)_4B_1(c)$=$4c(1-c)^4$ (5).
These are the probability of finding the Cu atoms with one Zn atom in their 4th nn positions {\it or} in the
2nd and the 3rd nn positions (4), and in their $j$(=1, 2, 3)th nn positions (5).

The relative intensity of the satellite Cu NQR spectrum of Y124 in Fig. 3(d)  
is smaller than that of Y123 in Fig. 3(c).
For Zn-substituted La$_{2-x}$Sr$_x$CuO$_4$, 
the disappearance of the Cu NQR signal without any satellite signal is observed~\cite{Yamagata1}.
Thus, it is likely that the observable satellite signal depends on the carrier-doping level
or the underlying magnetic correlation. 

As seen in Figs. 3(c) and 3(d), 
the Zn-doping dependence of the total intensity for Y123 ($x\leq$0.017) and Y124 ($x\leq$0.022) 
agrees with that of the case (3) within the experimental accuracy, 
where the up-to 3rd nn Cu sites to Zn are unobservable. 
The Zn doping dependence of the main intensity for Y123 ($x\leq$0.017) and Y124 ($x\leq$0.022) 
agrees with that of the case (2), where the up-to 4th nn Cu sites to Zn are unobservable. 
The Zn-nearest-neighbor Cu NQR spectrum will be shifted and broadened 
due to local crystalline or electric deformation, 
being unobservable probably due to a shorter relaxation time than a few $\mu$s. 
For Y123, the satellite signal agrees with either of the 4th nn Cu site 
or the 2nd and the 3rd nn Cu site to Zn [the case (4)]. 
With $x\leq$0.017, the former assignment is consistent with those to the total and the main signals, 
whereas with $x$=0.033, the latter is consistent with those to the total and the main ones.     
For Y124, the satellite signal agrees with the 1st, 2nd, or 3rd nn Cu site to Zn [the case (5)].
However, this is inconsistent with the up-to 3rd nn wipeout effect on the total intensity. 
Thus, the satellite resonance of Y124 may arise from the fraction of the 4th nn Cu site to Zn.

\subsubsection{``All-or-nothing" model}

 In the above analysis, one should note that the so-called ``all-or-nothing" model is implicitly assumed, 
where some nuclei are fully observable but the others are fully unobservable~\cite{Cohen}. 
This enables us to estimate the number of observed nuclei from the observed NQR signal intensity. 
In the actual sample, however,
there must be $T_2$ distribution. 
Figure 5 illustrates a two-pulse sequence with a Cu NQR spin-echo 
in a rotating frame with a Cu NQR frequency  
in the ``all-or-nothing" model (a) and in an actual system with $T_2$ distribution (b). 
In the case of Fig. 5(b), some of the observable nuclear spin-echoes after the two pulses will be underestimated  
due to a somewhat shorter $T_2$ than the totally corrected $T_2$ decay curve exp(-2$\tau/T_2$),
even when the Gaussian-times-exponential function empirically well describes $M(\tau)$.
Conversely, the $I_z$ fluctuation effect due to a short $T_1$~\cite{Curro} or 
the motional narrowing effect~\cite{Imai1,Imai2} on the transverse nuclear spin relaxation 
causes an exponential decay in a longer delay time $\tau$ but in a Gaussian decay in a shorter time $\tau$, 
then the total correction using a simple exponential function within the observable $\tau$ range will lead to overestimation of
the extrapolated $M(0)$. 
For both the main and the satellite signals, such a $T_2$ distribution or dynamical effect can cause 
overestimation or underestimation of the intensity.  
Thus, one should be careful with the results from intensity analysis. 
One can safely mention that the main signal does
not come from the Cu sites near Zn but that the satellite one from the Zn-neighbor Cu sites.

\begin{figure}
\epsfxsize=3.7in
\epsfbox{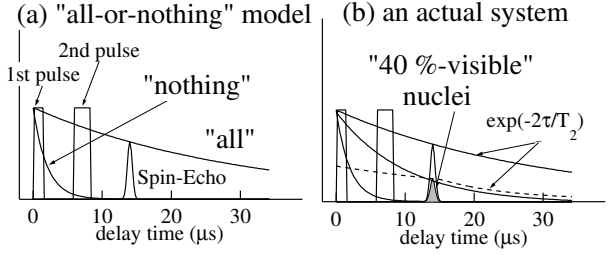}
\vspace{0.0cm}  
\caption{  
Illustrations of a pulse sequence of the first exciting $\pi$/2-pulse, 
the refocusing pulse $\pi$-pulse, 
and an NQR spin-echo in a rotating frame with an NQR frequency  
in the ``all-or-nothing" model (a) and in an actual system with $T_2$ distribution (b).
The ``40\%-visible" nuclei are supposed to have a shorter $T_2$ component with 60 \%
than the totally corrected $T_2$ decay curve. 
The solid and the dashed decay curves are transverse relaxation functions. 
}
\label{Pulse}
\end{figure}  

\subsubsection{Cu NQR frequency}
 
Here, we briefly mention the Cu NQR frequency.
The Cu NQR frequencies of stoichiometic cuprate oxides have been analyzed with several models
~\cite{eqQ}. 
For non-stoichiometic or impure compounds, 
although there is an attempt with numerical simulation~\cite{Imai}, it is more complicated, 
because local crystal strain ~\cite{ItohYSCO} or electric charge oscillation ~\cite{Haase} can affect
the line profile of Cu NQR spectrum.
Thus, we do not employ a specific model to analyze how the satellite Cu NQR frequency is shifted to the lower side  
than the main one. 
We just emphasize that 
Cu$^{2+}$ is a Jahn-Teller ion but Zn$^{2+}$ is a non-Jahn-Teller ion, 
so that the local crystalline distortion~\cite{Bridges} must modify the electric field gradient tensor at Zn-neighbor Cu sites,
differently from that at Cu sites away from Zn or 
from that in the pure Y123~\cite{Mali} or Y124~\cite{Zimmermann}. 
 
\subsection{Main and satellite $^{63}$Cu(2) nuclear spin-lattice relaxation}
\subsubsection{Zn-substitution effect on $^{63}$Cu(2) nuclear spin-lattice relaxation curves}
 
\begin{figure}
\epsfxsize=3.5in
\epsfbox{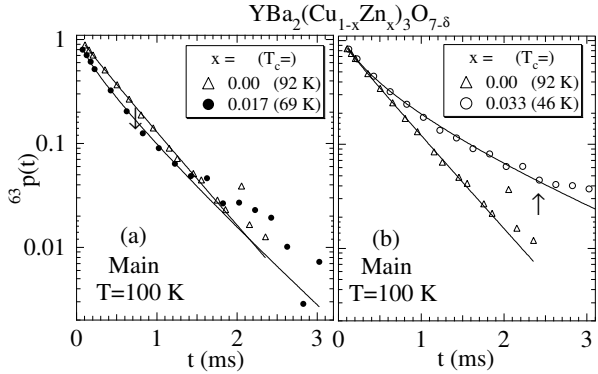}
\vspace{0.0cm}
\caption{
Zn-doping effect on the main $^{63}$Cu nuclear spin-echo recovery
curves $p(t)$ for Y123 with $x$=0.017 (a) and 0.033 (b) at $T$=100 K.  
The solid curves are the least-squares
fitting results using a theoretical function including a stretched exponential function (see the text).
}
\label{Recovery_vs_Zn}
\end{figure}

Figure 6 shows the main $^{63}$Cu(2) nuclear spin-lattice relaxation curve  
for Y123 with Zn concentrations $x$=0.017 (a) and 0.033 (b) at $T$=100 K. 
The result for Y124 has been reported in Ref.~\cite{Itoh1}. 
For Y123 with a small Zn concentration of $x$=0.017, 
the recovery curve $p(t)$ relaxes more quickly than that for pure Y123, 
which is consistent with the result for Y124 with Zn $x\leq$0.010. 
With further doping of $x$=0.033, $p(t >$ 1.0 ms) relaxes more slowly 
than that for pure Y123, 
which is similar to the result for Y124 with heavy doping of Zn $x$=0.022 at $T$=200 K~\cite{Itoh1}. 
 
\begin{figure}
\epsfxsize=3.5in
\epsfbox{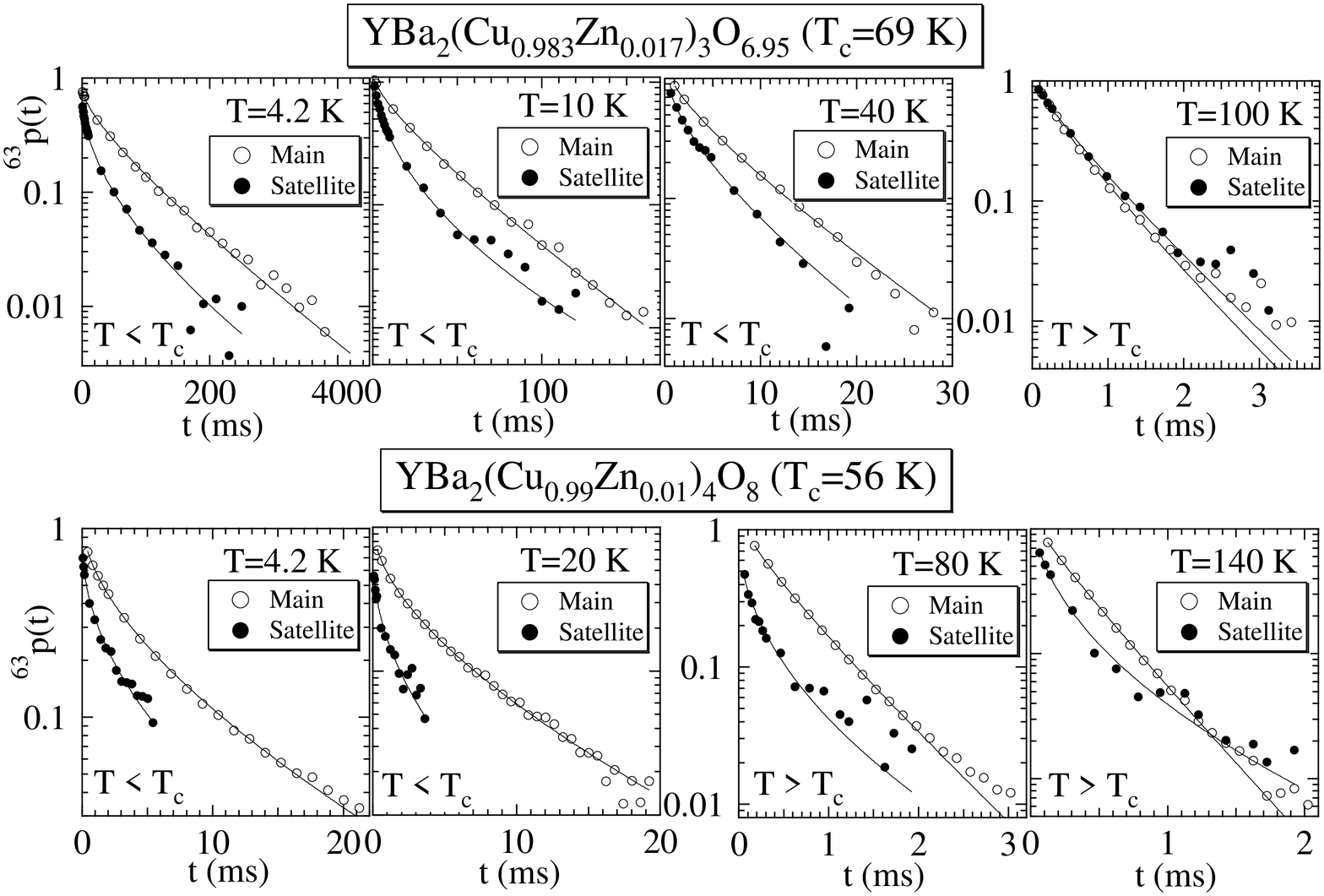}
\vspace{0.0cm}
\caption{
Temperature dependence of the $^{63}$Cu nuclear spin-echo recovery
curves $p(t)$ at the main and the satellite Cu NQR frequencies
for Y123 (upper) and Y124 (lower).  
The solid curves are the least-squares
fitting results using a theoretical function including a stretched exponential function (see the text).
}
\label{Recovery_vs_T}
\end{figure}
 
Figure 7 shows temperature dependence of the main and 
the satellite $^{63}$Cu nuclear spin-lattice relaxation curves $p(t)$
for Zn-substituted Y123 with $x$=0.017 (upper) and Y124 with $x$=0.01 (lower).
For Y123, the satellite signal recovers more quickly than the main one below $T_c$, but
they show nearly the same recovery curves above $T_c$. 
For Y124, however, both above and below $T_c$, the satellite signal recovers more quickly than the main one. 
With further Zn doping for Y124 ($x$=0.022), the difference between the main and the satellite signals tends to be reduced 
(not presented here), 
being consistent with the non-superconducting Y124~\cite{WilliamsCu}.
The result in Fig. 7 is evidence for different magnetic correlation
between the main and the satellite signals in the superconductors with small concentrations of Zn. 
   
\subsubsection{An analysis of nonexponential relaxation curves}

In Figs. 1, 2, 6, and 7, all the recovery curves are nonexponential functions. 
If the satellite signal comes from the same $j$th nn Cu sites by Zn and 
if the position dependence of Zn-induced relaxation process, i.e. 
the spatial distribution of Zn-induced moments or spin density, is
in axial symmetry along the $c$ axis, 
then the recovery curve must be a single exponential function. 
However, the actual curves are nonexponential ones. 
Thus, the satellite signals consist of non-single Cu sites, or 
the Zn-induced moment or spin density shows non-axial symmetry, e.g. $d$-wave symmetry.      

Nonexponential relaxation is frequently observed in the impure materials.
We have applied a conventional impurity-induced NMR relaxation theory~\cite{McHenry} to the impure high-$T_c$ cuprate
superconductors  and quantum spin systems~\cite{Itoh,Itoh1,Itoh2,Itoh3,Itoh4},  
which has been developed for dilute Heisenberg spin systems and dilute magnetic alloys. 
We believe that the analysis yields a vital, physical framework on the strongly correlated impure systems.   
The experimental $p(t)$ is analyzed 
by the exponential function times a stretched  exponential function $p(t)=p(0)\exp[-3t/(T_1)_{host}-\sqrt{3t/\tau_1}]$. 
$p(0)$, $(T_1)_{host}$ and $\tau_1$ are the fitting parameters. 
The numerical factor 3 is introduced to conform to the conventional expression of $T_1$~\cite{Moriya0}.
The solid and the dashed curves in Figs. 1, 2, 6, and 7 are the least-squares
fitting results using this approximate function.   
$(T_1)_{host}$ is the Cu nuclear spin-lattice  relaxation time 
due to the host Cu electron spin fluctuation via a hyperfine coupling. 
$\tau_1$ is the impurity-induced nuclear spin-lattice relaxation time, i.e., 
a spatial average of 1/$T_1(r)=A(r)^2S_{imp}(r, \omega_n)$ 
($r$ is a distance between a nuclear and an impurity-induced moment, 
$A(r)$ is a coupling constant between the nuclear spin and the impurity-induced moment, 
$S_{imp}(r, \omega_n)$ is the impurity-induced spin-spin correlation function~\cite{Kilian,Ohashi,Ohashi1}, 
and $\omega_n/2\pi$ is the nuclear resonance frequency)~\cite{McHenry,Moriya0}.
For an isolated local moment on the impurity site, 
one obtains 1/$T_1(r)\propto A(r)^2/T$ 
(a longitudinal direct dipole coupling $A(r)^2\propto 1/r^6$,  
or a two-dimensional RKKY interaction $A(r)^2\propto 1/r^4$)~\cite{McHenry}.
For an impurity-induced moment with a local coupling $A(r)\approx A(0)$,  
if $S_{imp}(r, \omega_n)\propto r^{-d}$ or e$^{-r/\xi}$, 
one obtains $p(t)\approx$ exp$[-(t/\tau_1)^{\alpha}]$ or exp(-ln$^2t$) 
($\xi$ is a correlation length, and $\alpha \propto d^{-1}$ is a constant)~\cite{Itoh2}.  

Instead of including all the direct or indirect nuclear-electron interactions,   
we assume a {\it minimal} model with a single impurity-induced relaxation 
and the host Cu electron spin relaxation. 
The recovery curve is expressed by $p(t)=p(0)\exp[-3t/(T_1)_{host}]\prod_i[(1-c)+c\exp(-3t/T_1({\bf r}_i))]$
($i$ is the Cu site index) as a function of
a lot of time constants of $(T_1)_{host}$, $T_1({\bf r}_1)$, $T_1({\bf r}_2)$, $T_1({\bf r}_3)$, $\cdots$.
For $c\ll$1, the product $\prod_i[\cdots]$ is approximated by $\approx\exp[-c\int d{\bf r}(1-\exp(-3t/T_1({\bf r})))]$, 
and then the spatial integral leads to the stretched exponential function with the single time constant $\tau_1$~\cite{McHenry}.
Eventually, only two fitting parameters on the time constants, i.e. $(T_1)_{host}$ and $\tau_1$ are obtained. 
The distribution of $T_1$ is taken into consideration through $T_1(r)$, 
although relations among $T_1(r)$, the $T_2$ distribution, and the broad Cu NQR frequency spectrum are not straightforward.   
	
In the actual fitting, we noticed that it is not easy to estimate the precise value of 1/$(T_1)_{host}$
in the case of $1/\tau_1\gg1/(T_1)_{host}$.
However, Zn-substitution effect on each relaxation time is fairly obvious.
The slow relaxation of $p(t >$ 1.0 ms) due to heavy doping of Zn as in Fig. 6(b) is the experimental fact. 
Within the present model, since $p(t)\rightarrow p(0)\exp[-3t/(T_1)_{host}]$ at $t\gg \tau_1/3$, then 
one estimates to decrease 1/$(T_1)_{host}$ with Zn doping of $x$=0.033. 
This estimation is not due to delicate fitting.          

\subsubsection{Zn-substitution effect on $^{63}\tau_1$ and $^{63}(T_1)_{host}$}

\begin{figure}
\epsfxsize=3.5in
\epsfbox{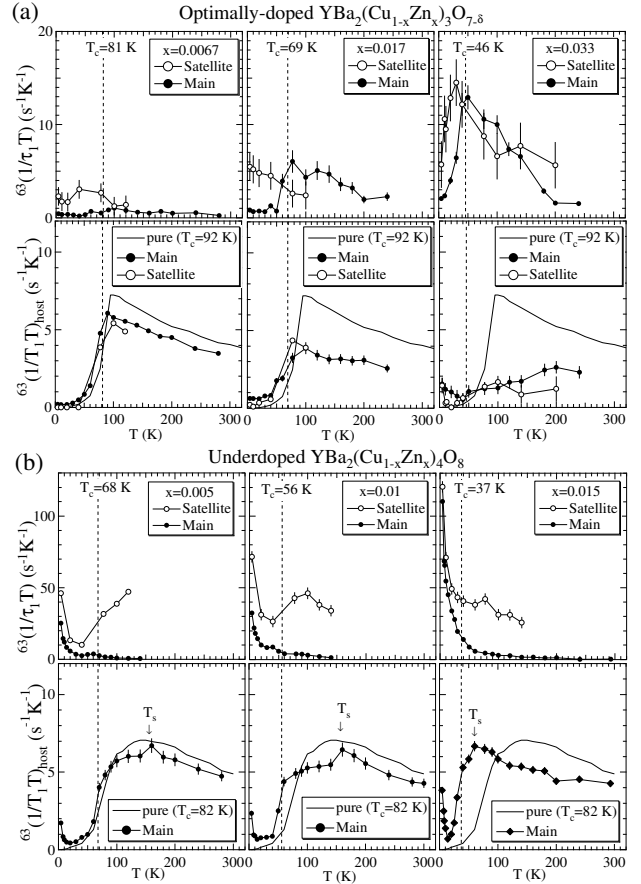}
\vspace{0.0cm}
\caption{
Zn-doping effect on the main and the satellite $^{63}$Cu(2) nuclear spin-lattice relaxation rates
$^{63}(1/\tau_1T$) and $^{63}(1/T_1T)_{host}$ 
as functions of temperature for Y123 (a) and for Y124 (b).  
The dashed lines indicate the respective $T_c$'s.
$T_s$ is the pseudo spin-gap temperature defined by the maximum temperature of $^{63}(1/T_1T)_{host}$.
}
\label{T1}
\end{figure} 

 Figure 8 shows the estimated $^{63}(1/\tau_1T)$ and $^{63}(1/T_1T)_{host}$ 
of the satellite and of the main signals as functions of
temperature for Zn-substituted Y123 (a) and Y124 (b). 
First, the satellite $^{63}(1/\tau_1T)$ for Y124 at $T$=4.2 K is an order of magnitude higher than that for Y123. 
Second, the temperature dependence of $^{63}(1/\tau_1T)$ of the satellite signal is different between
Y123 and Y124.
For Y124 at lower temperatures below $T_c$, $^{63}(1/\tau_1T)$ increases as temperature is decreased, 
in contrast to $T$ independent or slight decrease of $^{63}(1/\tau_1T)$ for Y123. 
Third, the difference in the $T$ dependence of $^{63}(1/\tau_1T)$ between the main and the satellite signals
is marked below $T_c$ for Y123 but above $T_c$ for Y124.  
Fourth, the main $^{63}(1/T_1T)_{host}$ above about 100 K decreases with Zn doping for both systems.
Fifth, the satellite $^{63}(1/T_1T)_{host}$ was estimated to be $\ll$ 1.0 s$^{-1}$K$^{-1}$ for Y124 with Zn. 
From these results, one can see that the local magnetic correlation around Zn is 
different between underdoped and optimally doped samples. 

The Cu(2) nuclear spin-lattice relaxation time probes 
the in-plane antiferromagnetic dynamical spin susceptibility at an NMR/NQR frequency
~\cite{Shastry,MMP,MTU,Bulut}
and homogeneity of the CuO$_2$ plane in real space~\cite{McHenry,Moriya0}. 
Relation between our results on the Cu NQR relaxation and the previous $^{89}$Y NMR results~\cite{Mahajan}  
is not straightforward.  

$\it ^{63}(1/\tau_1T)$: 
The obvious difference in $^{63}(1/\tau_1T)$ between the main and the satellite signals indicates 
an evidence for the Zn-induced virtual bound state~\cite{Balatsky,Onishi,Salkola} 
through the locally enhanced magnetic correlation~\cite{Ohashi,Ohashi1,Bulut1,Poilblanc}.    
For underdoped Y124, the difference in $^{63}(1/\tau_1T)$ may indicate 
that Zn-induced ``localized moments" persist both above and below $T_c$~\cite{Bobroff}. 
For optimally doped Y123, the below-$T_c$ difference in $^{63}(1/\tau_1T)$ may indicate   
that the Zn-induced, generalized RKKY oscillation is extended around Zn below $T_c$ but localized above $T_c$.
This is consistent with a prediction in the perturbative mode-mode coupling theory~\cite{Ohashi1,Bulut1},  
where the spatially extended potential scattering by Zn plays a key role~\cite{Balatsky1,Xiang}.   
The separated frequency of the satellite Cu NQR spectrum may also support such a finite range effect
of Zn impurity potential. 
According to Ref.~\cite{Ohashi1}, 
the 1st nn potential term $\left|V_2\right|$ will be larger in Y124 than Y123, 
because the satellite $^{63}(1/\tau_1T)$ is larger than the main one in Y124 above and below $T_c$.   
The local electronic state induced by Zn reflects 
the matrix antiferromagnetic correlation~\cite{Ohashi,Ohashi1,Bulut1,Poilblanc}. 
The optimally doped Y123 still has the underlying antiferromagnetic correlation~\cite{Imai0,Imai3}.   

$\it ^{63}(1/T_1T)_{host}$: 
For Y124, the absence or a small value of the satellite $^{63}(1/T_1T)_{host}$ suggests that
the conduction electrons in the underdoped system are expelled from the vicinity of Zn atoms, or 
that the pseudo spin-gap effect is enhanced near Zn. 
A remnant of the pseudo spin-gap effect can be seen as a decrease of $^{63}(1/\tau_1T)$ below $T$=100 K 
with $x$=0.005 and 0.01 in Fig. 8(b). 
For Y123, nearly the same $^{63}(1/T_1T)_{host}$ between the satellite and the main signals indicates 
that the conduction electrons uniformly distribute near and away from Zn both above and below $T_c$, 
in contrast to the underdoped Y124. 
The pure Y124 is a stoichiometric, homogeneous, underdoped system.
Nevertheless, it has some inherent electronic instability, e.g. magnetic instability, 
which may lead to an inhomogeneous magnetic response. 
The underdoped electronic state of Y124 is located more close to the magnetic instability
than the optimally doped one of Y123.  

The above-$T_c$ decrease in $^{63}(1/T_1T)_{host}$ with Zn doping for both systems 
indicates that Zn suppresses the host antiferromagnetic spin susceptibility~\cite{Itoh3}. 
This is recently reproduced by a numerical quantum Monte Carlo simulation with the dynamical
cluster approximation based on the two-dimensional $t-J$ model~\cite{Jarrell}.       

\section{Conclusion}
\label{sec:conclusion}

	We succeeded in measuring the Zn-induced satellite Cu(2) NQR nuclear spin-lattice relaxation curves
in the superconductors Y123 and Y124 above and below $T_c$ in a zero external magnetic field. 
With doping a small amount of Zn, the electron spin correlation near Zn is enhanced in the optimally
doped system as well as in the underdoped one. 
The optimally doped Y123  
still has an underlying antiferromagnetic correlation.  
The local electronic state induced by Zn reflects it. 

\acknowledgments

	We would like to thank M. Matsumura, H. Yamagata (YBa$_2$Cu$_3$O$_{6.89}$) and 
Y. Ueda (YBa$_2$Cu$_3$O$_{6.98}$) for providing their samples and for helpful discussions, 
and Y. Ohashi, M. Ogata, M. H. Julien, and K. Ishida for stimulating discussions.   
This work was supported by the New Energy and Industrial 
Technology Development Organization (NEDO) as Collaborative Research and 
Development of Fundamental Technologies for Superconductivity Applications.  


\end{document}